\documentclass[pra,twocolumn,superscriptaddress,showpacs,preprintnumbers,nofootinbib]{revtex4-1}
\usepackage{amsmath,bm,graphicx,hyperref}

\renewcommand\d{\partial}
\newcommand\+{\dagger}
\newcommand\<{\langle}
\renewcommand\>{\rangle}
\newcommand\x{\bm{x}}
\newcommand\X{\bm{X}}
\renewcommand\P{\bm{P}}

\renewcommand\Re{\mathrm{Re}}
\newcommand\arcsinh{\mathrm{arcsinh}}
\newcommand\C{\mathcal{C}}
\newcommand\K{\mathcal{K}}
\renewcommand\O{\mathcal{O}}

\begin{document}
\preprint{LA-UR-12-10338}

\title{Weakly bound molecules trapped with discrete scaling symmetries}

\author{Yusuke~Nishida}
\affiliation{Theoretical Division, Los Alamos National Laboratory,
Los Alamos, New Mexico 87545, USA}
\author{Dean~Lee}
\affiliation{Department of Physics, North Carolina State University,
Raleigh, North Carolina 27695, USA}

\begin{abstract}
 When the scattering length is proportional to the distance from the
 center of the system, two particles are shown to be trapped about the
 center.  Furthermore, their spectrum exhibits discrete scale
 invariance, whose scale factor is controlled by the slope of the
 scattering length.  While this resembles the Efimov effect, our system
 has a number of advantages when realized with ultracold atoms.  We also
 elucidate how the emergent discrete scaling symmetry is violated for
 more than two bosons, which may shed new light on Efimov physics.  Our
 system thus serves as a tunable model system to investigate universal
 physics involving scale invariance, quantum anomaly, and
 renormalization group limit cycle, which are important in a broad range
 of quantum physics.
\end{abstract}

\date{February 2012}


\maketitle

\section{Introduction}
When particles attract by a short-range interaction with a large
scattering length, their low-energy physics becomes {\em universal\/},
i.e., independent of microscopic details~\cite{Braaten:2006}.  Ultracold
atoms are ideal to study such universal physics because of the
tunability of interatomic interactions and can provide insights
applicable in a broad range of physics.  One of the most striking
phenomena in universal systems is the Efimov effect, i.e., the formation
of an infinite tower of three-body bound states characterized by
discrete scale invariance~\cite{Efimov:1970}.  Although the Efimov
effect was originally predicted in the context of nuclear physics, it is
now subject to extensive research in ultracold
atoms~\cite{Ferlaino:2010}.

In this paper, we propose novel systems in which {\em two\/} particles
exhibit discrete scale invariance in their spectrum.  In order for this
to happen, their interaction needs to be scale
invariant~\cite{Nishida:2011}.  It is usually considered that the
short-range interaction can be scale invariant only when the scattering
length $a$ is set to be zero or infinite.  However, there is another
possibility.  The scattering length is made space dependent and tuned to
be proportional to the distance from the center of the
system:\footnote{Here, translational symmetries are sacrificed.  A
similar but different way to achieve the scale-invariant interaction is
to vary the scattering length only in a certain direction, for example,
$a(\x)=c|z|$.  In this case, both the translational and rotational
symmetries are partially broken.  While we expect similar physics, we
shall not deal with such cases in this paper.}
\begin{equation}
 a(\x) = c|\x|.
\end{equation}
This interaction is {\em scale invariant\/} because there is no
dimensional parameter in it.

The emergence of the discrete scale invariance for $c>0$ can be
understood intuitively by using the Born-Oppenheimer approximation.
Suppose one particle is much heavier than the other particle.  With the
heavy particle fixed at $\x$, the light particle forms a bound state
with binding energy $-\hbar^2/[2\mu a(\x)^2]$, which in turn acts as an
effective potential for the heavy particle.  Therefore, one can design
any attractive potential by tuning the space dependence of the
scattering length.  In particular, when $a(\x)=c|\x|$, the effective
potential becomes an inverse square potential for which it is well known
that the spectrum exhibits discrete scale
invariance~\cite{Landau-Lifshitz}.  Since two particles are trapped
about the center of the system with discrete scaling symmetry, we shall
refer to our system as a {\em scaling trap\/}.

This conclusion can be established for any mass ratio by solving the
two-body problem exactly with the space-dependent scattering length.
While our idea works in any spatial dimensions, we shall give extensive
and detailed analyses in one dimension
(Secs.~\ref{sec:two-body}--\ref{sec:EFT}) and then present key results
in two and three dimensions (Sec.~\ref{sec:2D-3D}).  Remarks on
experimental realization with ultracold atoms are given in
Sec.~\ref{sec:remarks}.

\section{Two particles in one dimension \label{sec:two-body}}

\subsection{Bound-state solutions}
Two interacting particles in one dimension are described by the
Schr\"odinger equation (hereafter $\hbar=1$):
\begin{equation}\label{eq:schrodinger}
 \left[-\frac{\nabla_X^2}{2M}-\frac{\nabla_x^2}{2\mu}+V(X,x)\right]
  \psi(X,x) = E\,\psi(X,x).
\end{equation}
Here, $M=m_1+m_2$ and $\mu=m_1m_2/(m_1+m_2)$ are the total and reduced
masses and $X=(m_1x_1+m_2x_2)/M$ and $x=x_1-x_2$ are the center-of-mass
and relative coordinates, respectively.  For a zero-range interaction
whose strength depends on the position $X$, the interaction potential is
written as
\begin{equation}
 V(X,x) = -\frac1{\mu a(X)}\delta(x),
\end{equation}
where $a(X)$ is the space-dependent scattering length.  For a
bound-state solution with $E\equiv-\kappa^2/(2M)<0$, the Schr\"odinger
equation is formally solved by
\begin{equation}\label{eq:wave_function}
 \tilde\psi(P,p) = \frac1{\frac{P^2}{2M}+\frac{p^2}{2\mu}+\frac{\kappa^2}{2M}}
  \frac1\mu\int\!\frac{dP'}{2\pi}\frac1{\tilde a(P{-}P')}\tilde\chi(P'),
\end{equation}
where $\tilde\psi(P,p)\equiv\int\!dXdx\,e^{-iPX-ipx}\,\psi(X,x)$ is the
wave function in momentum space and
$\frac1{\tilde a(P)}\equiv\int\!dX\,e^{-iPX}\frac1{a(X)}$ is the Fourier
transform of the inverse scattering length.  By integrating both sides
of Eq.~(\ref{eq:wave_function}) over $p$, we obtain an integral equation
solved by $\tilde\chi(P)\equiv\int\!\frac{dp}{2\pi}\tilde\psi(P,p)$:
\begin{equation}\label{eq:integral_eq}
 \tilde\chi(P) = \frac1{\sqrt{P^2+\kappa^2}}\sqrt\frac{M}{\mu}
  \int\!\frac{dP'}{2\pi}\frac1{\tilde a(P{-}P')}\tilde\chi(P').
\end{equation}
We note that when $a(X)=a>0$ is uniform, there is a single bound state
with binding energy $E=-1/(2\mu a^2)$.

Now for the linearly space-dependent scattering length
$1/a(X)=1/(c|X|)$, $1/\tilde a(P)$ is ill defined because of the
divergence at $X=0$.  This divergence needs to be regularized, for
example, by a sharp cutoff $1/a(X)=\theta(|X|-\epsilon)/(c|X|)$ or by a
smooth cutoff $1/a(X)=1/(c\sqrt{X^2+\epsilon^2})$.  In either case,
the limit of an infinitesimal cutoff $\epsilon\to0$ leads to
$1/\tilde a(P)\to-(2/c)\ln(\epsilon|P|)$ for which the analytic solution
of the integral equation (\ref{eq:integral_eq}) is found to be
\begin{subequations}\label{eq:solution}
 \begin{align}
  \tilde\chi_+(P) &= N_+\frac{\cos[s_+\arcsinh(P/\kappa)]}{\sqrt{(P/\kappa)^2+1}}, \\[8pt]
  \tilde\chi_-(P) &= N_-\frac{\sin[s_-\arcsinh(P/\kappa)]}{i\sqrt{(P/\kappa)^2+1}}.
 \end{align}
\end{subequations}
The $+$ ($-$) sign corresponds to the even (odd) parity, and
\begin{equation}
 |N_\pm|^2 \equiv \sqrt{\frac{\mu}{M}}
  \frac{2\pi}{1\pm\frac{\pi s_\pm}{\sinh\pi s_\pm}}
\end{equation}
is the normalization constant to ensure
$\int\!\frac{dPdp}{(2\pi)^2}|\tilde\psi_\pm(P,p)|^2=1$ and $s_\pm>0$
solves
\begin{equation}\label{eq:s_1D}
 1 = \frac1c\sqrt\frac{M}{\mu}\frac{\coth\frac{\pi s_+}2}{s_+}, \qquad
  1 = \frac1c\sqrt\frac{M}{\mu}\frac{\tanh\frac{\pi s_-}2}{s_-}.
\end{equation}
Details of deriving Eqs.~(\ref{eq:solution}) and (\ref{eq:s_1D}) are
presented in the Appendix.  Note that $s_+$ has a solution for any
$c>0$ while $s_-$ has a solution only for $0<c<\pi\sqrt\frac{M}{4\mu}$.
The latter range of $c$ is assumed below unless otherwise stated.

\begin{figure}[t]
 \includegraphics[width=0.9\columnwidth,clip]{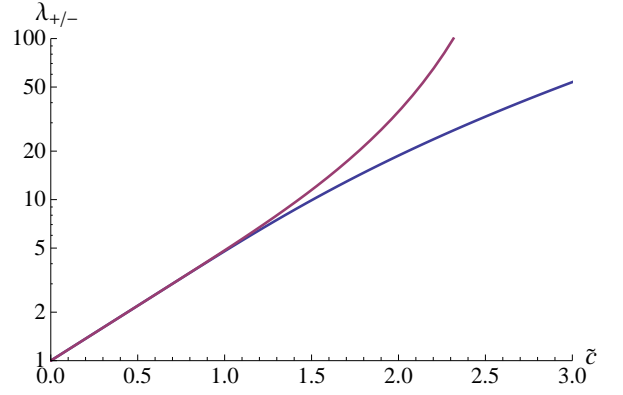}
 \caption{(Color online) Scale factors $\lambda_+=e^{\pi/s_+}$ (lower
 curve) and $\lambda_-=e^{\pi/s_-}$ (upper curve) determined by
 Eq.~(\ref{eq:s_1D}) as functions of the slope of the scattering length
 $\tilde{c}\equiv c\sqrt{4\mu/M}$.  \label{fig:scale-factor_1D}}
\end{figure}

Then the inverse Fourier transform of $\tilde\chi_\pm(P)$ leads to the
wave function with two particles at the same point,
$\chi_\pm(X)=\psi_\pm(X,0)$, where
\begin{subequations}\label{eq:wave_function2}
 \begin{align}
  \psi_+(X,0) &= \kappa\,N_+\frac{\cosh\frac{\pi s_+}2}{\pi}
  K_{is_+}(\kappa|X|), \\[8pt]
  \psi_-(X,0) &= \kappa\,N_-\frac{\sinh\frac{\pi s_-}2}{\pi}
  K_{is_-}(\kappa|X|)\ \mathrm{sgn}(X).
 \end{align}
\end{subequations}
An important observation is that this wave function toward the origin
$X\to0$ oscillates as
$\psi_\pm(X,0)\propto K_{is_\pm}(\kappa|X|)\to|\Gamma(is_\pm)|\cos[s_\pm\ln(\kappa|X|/2)-\arg\Gamma(is_\pm)]$.
The phase of this oscillation is fixed by the precise behavior of $a(X)$
near $X=0$, which is not universal and depends on experimental setups.
However, what is universal is that because of the logarithmic
periodicity in $\kappa$, if $\kappa=\kappa_\pm$ is a solution, then
$\kappa=e^{-\pi n/s_\pm}\kappa_\pm$ are all solutions.  Therefore, in
each parity channel, there exists an infinite tower of two-body bound
states characterized by discrete scale invariance:
\begin{equation}\label{eq:spectrum}
 E_\pm^{(n)} = -e^{-2\pi n/s_\pm}\,\frac{\kappa_\pm^2}{2M}. 
\end{equation}
This is exactly the same physics as the Efimov effect while the
difference should be emphasized that our bound state consisting of two
particles is trapped about the center of the system (see
Fig.~\ref{fig:density} below) and the scale factor
$\lambda_\pm\equiv e^{\pi/s_\pm}$ is tunable by the slope of the
scattering length as shown in Fig.~\ref{fig:scale-factor_1D}.  We also
note that the full scale invariance demonstrated by the classical
Hamiltonian (\ref{eq:schrodinger}) with $a(X)=c|X|$ is broken down to
its discrete subset by the scale $\kappa_\pm$ generated in quantum
mechanics.  This is known as a {\em quantum
anomaly\/}~\cite{Holstein:1993}.

\begin{figure*}[t]\hfill
 \includegraphics[width=0.9\columnwidth,clip]{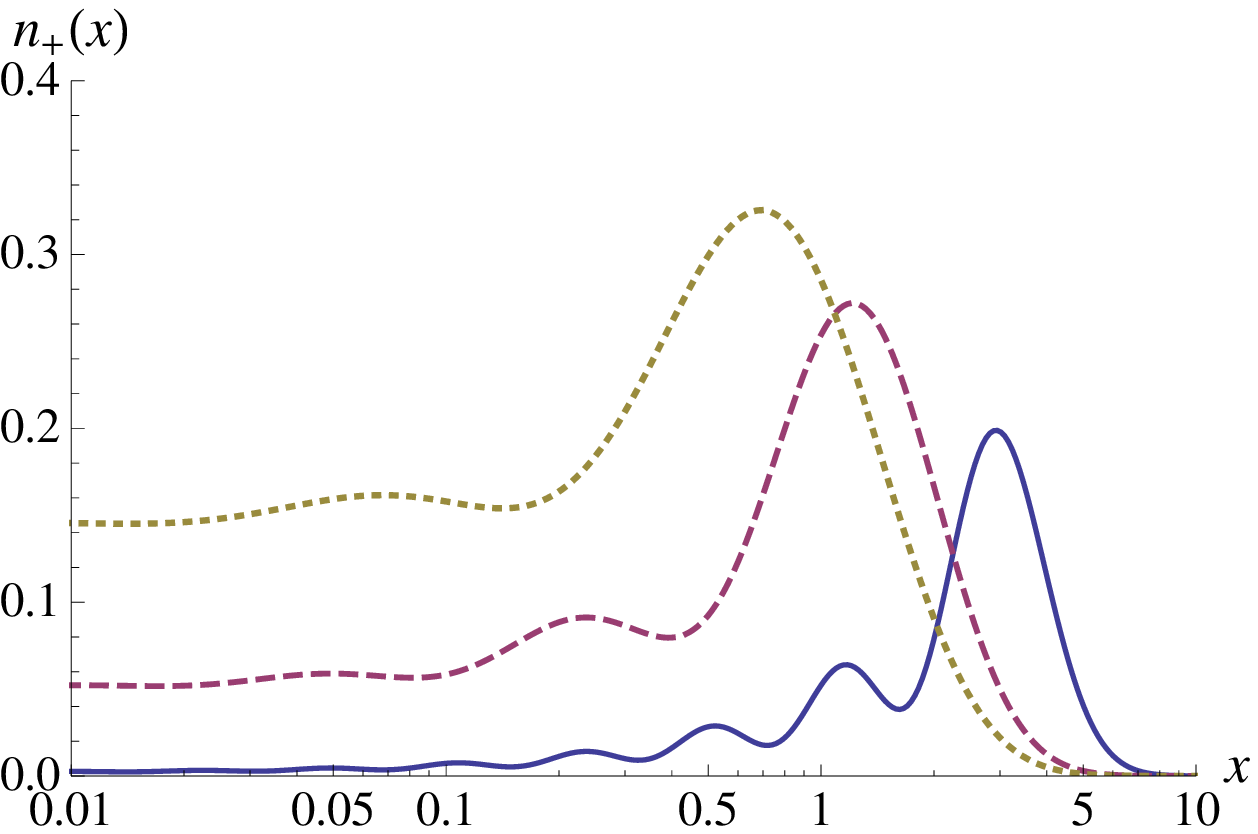}\hfill\hfill
 \includegraphics[width=0.9\columnwidth,clip]{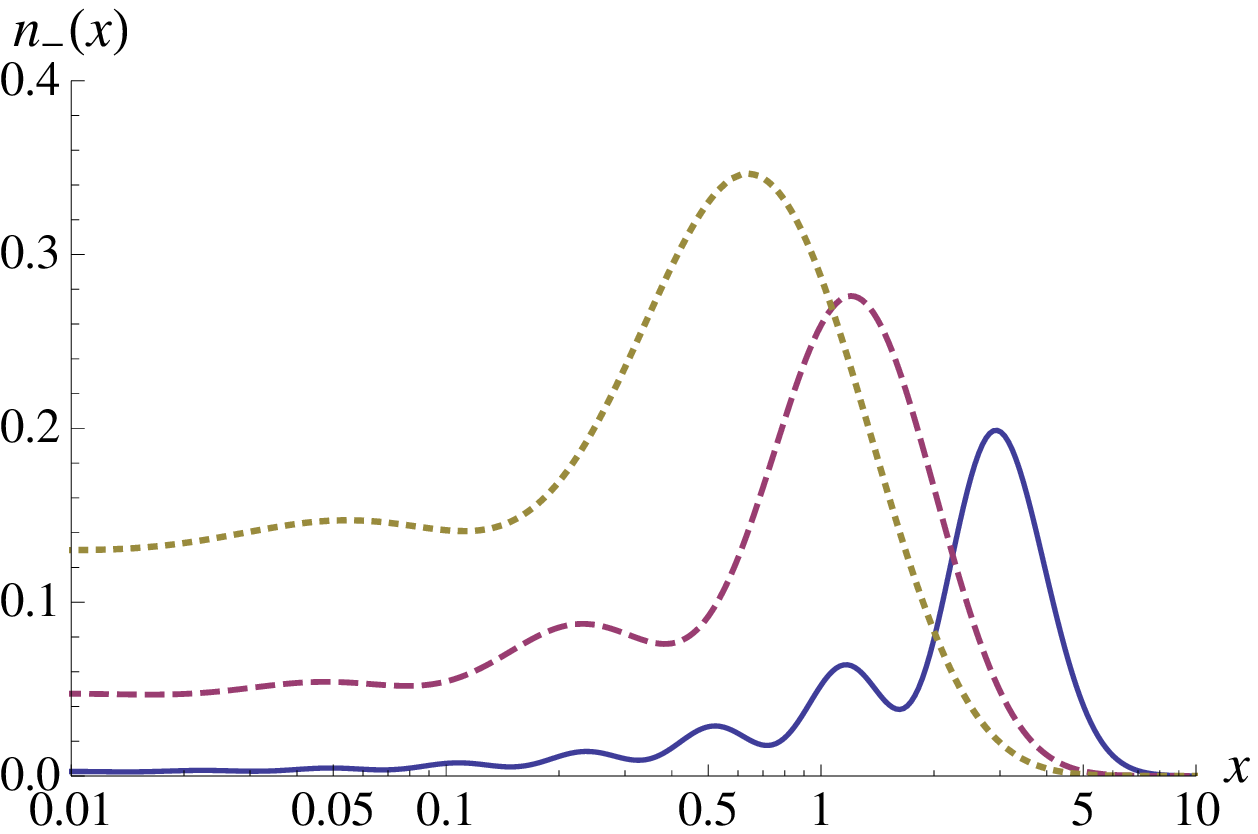}\hfill\hfill
 \caption{(Color online) Single-particle density distribution
 $n_\pm(x_1)$ in the two-particle trapped state with even (left) or odd
 parity (right panel) in units of $\kappa_\pm=1$.  Solid, dashed, and
 dotted curves correspond to $c=0.5,\,1,\,1.5$, respectively, with equal
 masses $m_1=m_2$.  \label{fig:density}}
\end{figure*}

\begin{figure*}[t]\hfill
 \includegraphics[width=0.9\columnwidth,clip]{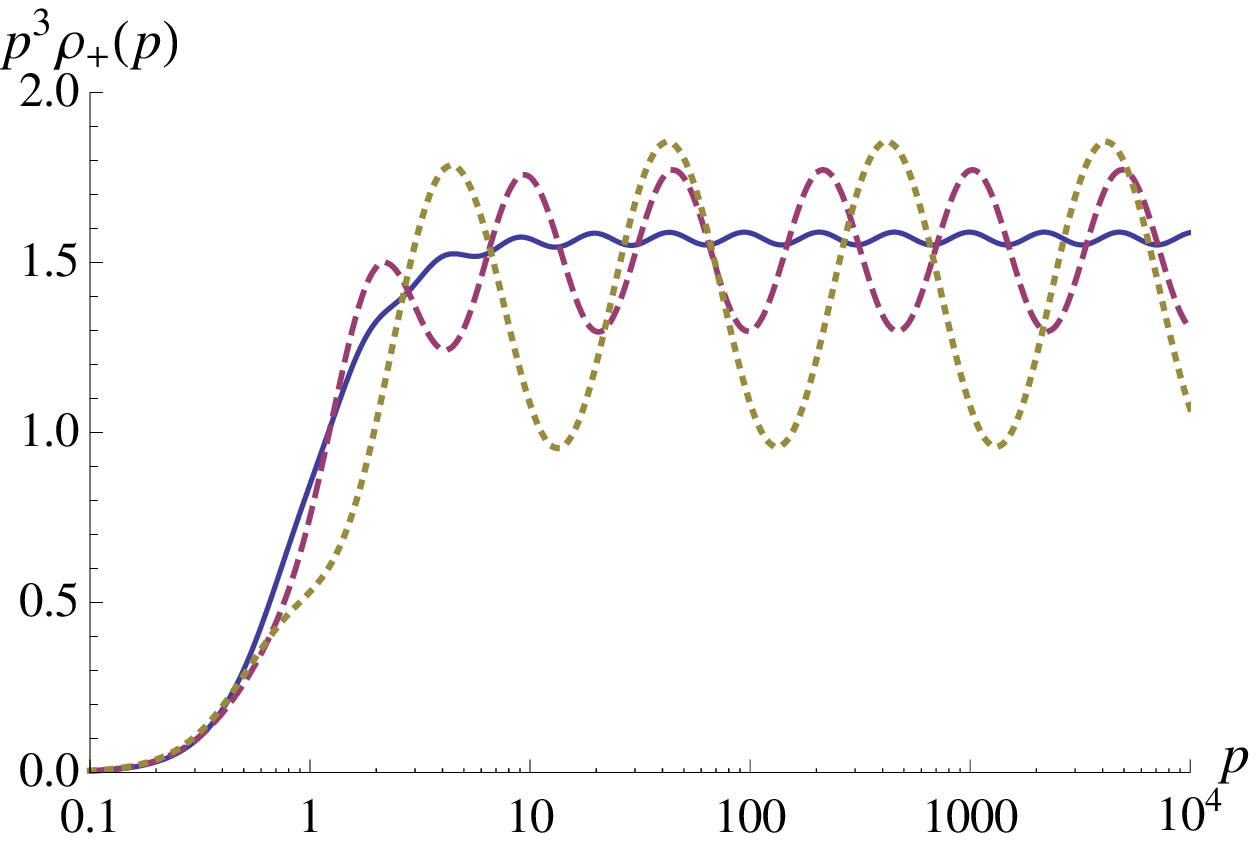}\hfill\hfill
 \includegraphics[width=0.9\columnwidth,clip]{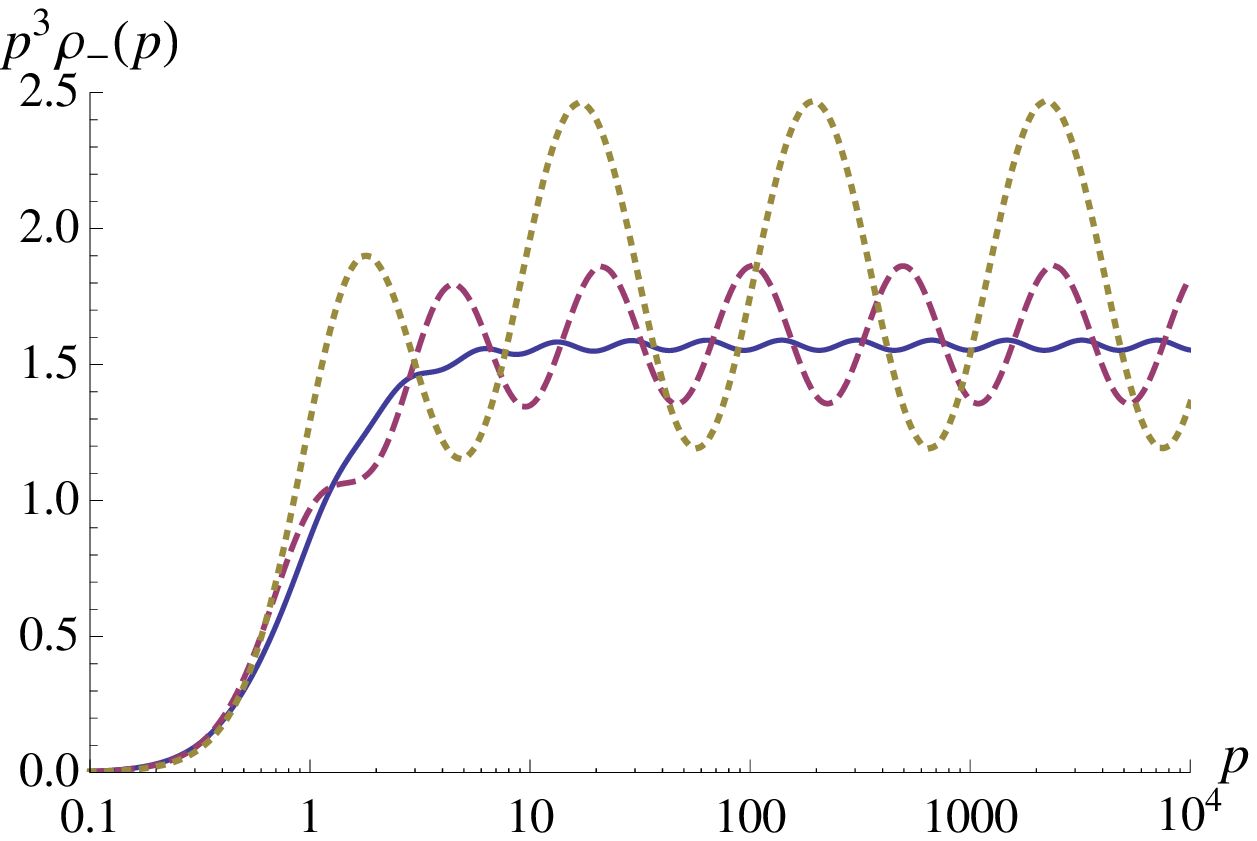}\hfill\hfill
 \caption{(Color online) Single-particle momentum distribution
 $p_1^3\rho_\pm(p_1)$ in the two-particle trapped state with even (left)
 or odd parity (right panel) in units of $\kappa_\pm=1$.  Solid, dashed,
 and dotted curves correspond to $c=0.5,\,1,\,1.5$, respectively, with
 equal masses $m_1=m_2$.  \label{fig:momentum}}
\end{figure*}

\subsection{Density and momentum distributions}
We now determine single-particle density and momentum distributions in
the two-particle trapped state with binding energy
$E=-\kappa_\pm^2/(2M)$.  With the use of the wave function obtained from
Eqs.~(\ref{eq:wave_function})--(\ref{eq:solution}),
\begin{equation}
 \tilde\psi_\pm(P,p) = \frac{2\sqrt{\frac{\mu}{M}}\sqrt{P^2+\kappa_\pm^2}}
  {p^2+\frac{\mu}{M}\!\left(P^2+\kappa_\pm^2\right)}\tilde\chi_\pm(P),
\end{equation}
the density distribution of a particle with mass $m_1$,
$n_\pm(x_1)\equiv\int\!dx\bigl|\psi_\pm\!\left(x_1{-}\frac{m_2}{M}x,x\right)\bigr|^2$,
and its momentum distribution
$\rho_\pm(p_1)\equiv\int\!\frac{dP}{2\pi}\bigl|\tilde\psi_\pm\!\left(P,p_1{-}\frac{m_1}{M}P\right)\bigr|^2$
are plotted in Figs.~\ref{fig:density} and \ref{fig:momentum},
respectively, for $c=0.5,\,1,\,1.5$ with equal masses $m_1=m_2$.  In
particular, the momentum distribution at $|p_1|/\kappa_\pm\to\infty$ has
an oscillatory large momentum tail
$\rho_\pm(p_1)\to\kappa_\pm^2t_\pm(p_1)$ given by
\begin{equation}\label{eq:tail}
 \begin{split}
  t_\pm(p_1) &\equiv \frac{|N_\pm|^2}{|p_1|^3}\Biggl\{\frac12
  \pm \Re\Biggl[\left(2\sqrt2\,\frac{|p_1|}{\kappa_\pm}\right)^{2is_\pm} \\[4pt]
  &\quad \times \frac{(1-is_\pm)\cosh\frac{\pi s_\pm}2+s_\pm\sinh\frac{\pi s_\pm}2}
  {2\cosh\pi s_\pm}\Biggr]\Biggr\}.
 \end{split}
\end{equation}
This logarithmic oscillation signals the discrete scale invariance, and
exactly the same tail emerges in any few-body or many-body state as we
will show later.  In contrast to the Efimov effect in which the
oscillatory tail appears at the subleading
order~\cite{Castin:2011,Braaten:2011}, it appears at the leading order
in our scaling trap.  This will make its observation easier by a
time-of-flight measurement in ultracold-atom experiments.

\section{More than two particles}

\subsection{Identical bosons}
A longstanding problem in Efimov physics is whether the discrete scale
invariance demonstrated for three particles persists for more
particles~\cite{Braaten:2006,Ferlaino:2010}.  It has been established
that this is the case for four bosons~\cite{Deltuva:2012} while the
situation is less clear for larger numbers of particles.  Here, we
elucidate that the scaling trap realizes a novel pattern of
discrete-scaling-symmetry violation for bosons.  Different particle
sectors obey different scaling laws, and incommensurate scalings among
them result in the breakdown of discrete scale invariance.

This pattern can be explained easily in the limit $c\ll1$ where the
Born-Oppenheimer approximation is applicable.  Recall that
one-dimensional bosons with a zero-range attraction form an $N$-body
bound state with binding energy $-N(N^2-1)/(6ma^2)$~\cite{Takahashi}.
For the linearly space-dependent scattering length $a=c|X|$, this
binding energy acts as an effective potential for the center-of-mass
motion of the $N$-body cluster, which is described by
\begin{equation}
 \left[-\frac{\nabla_X^2}{2Nm}-\frac{N(N^2-1)}{6mc^2X^2}\right]
  \psi_N(X) = E\,\psi_N(X).
\end{equation}
Since this is the inverse square potential, $N$ bosons form an infinite
tower of trapped states characterized by a discrete scaling symmetry set
by a scale factor $\lambda_N=e^{\pi/s_N}$ with
\begin{equation}
 s_N = \frac1c\sqrt{\frac{N^2(N^2-1)}3} + O(c^0).
\end{equation}
However, these $N$-body trapped states for $N\geq3$ are actually
unstable resonances coupled with continuum states.  For example, when
$N=3$, there are continuum states composed of a free particle and a
two-body trapped state obeying a discrete scaling symmetry set by
$\lambda_2$.  Because dilatations with respect to
$\lambda_2\approx e^{\pi c/2}$ and
$\lambda_3\approx e^{\pi c/(2\sqrt6)}$ are incommensurate, the discrete
scale invariance breaks down for three bosons and hence more.

Further insights into the nature of $N$-boson resonances can be
developed in the same limit $c\ll1$.  Because the local binding energy
scales as $-1/(c|X|)^2$, their relative wave function is localized
within a separation $\sim c|X|$.  On the other hand, their
center-of-mass wave function is $\psi_N(X)\sim K_{is_N}(\kappa_N|X|)$,
which oscillates rapidly at $\kappa_N|X|\ll1$ and decays exponentially
at $\kappa_N|X|\gg1$.  This $N$-body cluster can decay into a deeper
$N'$-body cluster with $2\leq N'<N$, and its wave function also
oscillates rapidly but with a different logarithmic period
$\pi/s_{N'}>\pi/s_N$.  Because of the resulting small overlap between
their wave functions, the $N$-boson resonances are expected to have
small decay widths and thus obey an {\em approximate\/} discrete scaling
law set by $\lambda_N$.

\begin{figure}[t]
 \includegraphics[width=0.9\columnwidth,clip]{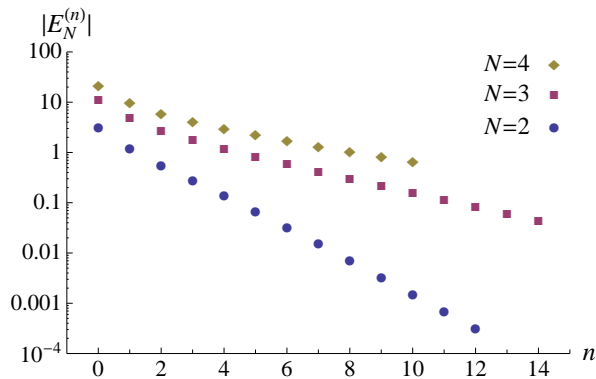}
 \caption{(Color online) Lattice results for $N$-boson resonance energy
 $E_N^{(n)}$ in units of $\mu=1$ and unit lattice spacing versus
 excitation number $n$ at $c=0.25$.  \label{fig:resonances}}
\end{figure}

To confirm this consideration, we numerically computed $N$-boson
resonance energies for $N=2,\,3,\,4$.  Here, a Hamiltonian lattice
formalism and iterative eigenvector methods were used in a semi-infinite
system with a hard-wall boundary at the origin.\footnote{While this
one-sided system, in general, does not have the same spectrum as the
two-sided system studied in the rest of this paper, they show the same
scaling behaviors in the limit $c\ll1$ where the Born-Oppenheimer
approximation is applicable.}
At $c=0.25$, decay widths of resonances are indeed found to be
negligible, and scaling behaviors of resonance energies are clearly seen
in Fig.~\ref{fig:resonances} as linear behaviors for large $n$ in the
logarithmic plot.  Their scale factors are extracted as
$\lambda_2\approx1.482(2)$, $\lambda_3\approx1.172(2)$, and
$\lambda_4\approx1.10(2)$, which are in good agreement with the
Born-Oppenheimer approximation
\begin{equation}
 \lambda_N \approx e^{\pi c/\sqrt{N^2(N^2-1)/3}}
  \approx 1.481,\ 1.174,\ 1.092,
\end{equation}
for $N=2,\,3,\,4$, respectively.

\subsection{Two-component fermions}
In contrast to bosons, we did not observe any resonances in the spectrum
of three fermions (two of one component and one of the other) with equal
masses.  This is indeed expected because one-dimensional two-component
fermions with equal masses and with a zero-range attraction do not form
any bound states with more than two fermions~\cite{Takahashi}.
Therefore, their discrete scaling symmetry cannot be violated by the
pattern elucidated above.

\begin{figure}[t]
 \includegraphics[width=0.9\columnwidth,clip]{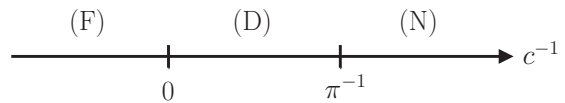}
 \caption{Expected three phases with (F) full, (D) discrete, and (N) no
 scaling symmetries for two-component fermions with equal masses as a
 function of the inverse slope $c^{-1}$.  \label{fig:phase_diagram}}
\end{figure}

\begin{figure*}[t]\hfill
 \includegraphics[width=0.9\columnwidth,clip]{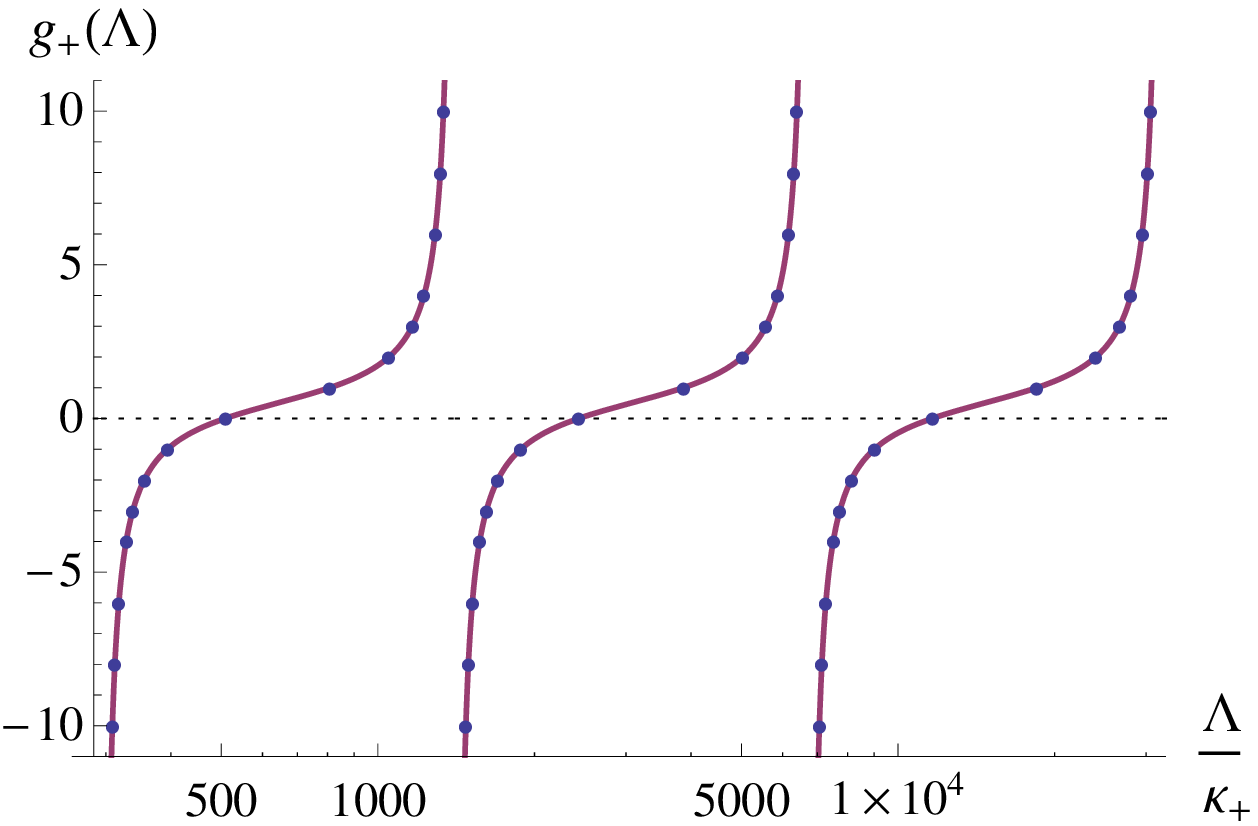}\hfill\hfill
 \includegraphics[width=0.9\columnwidth,clip]{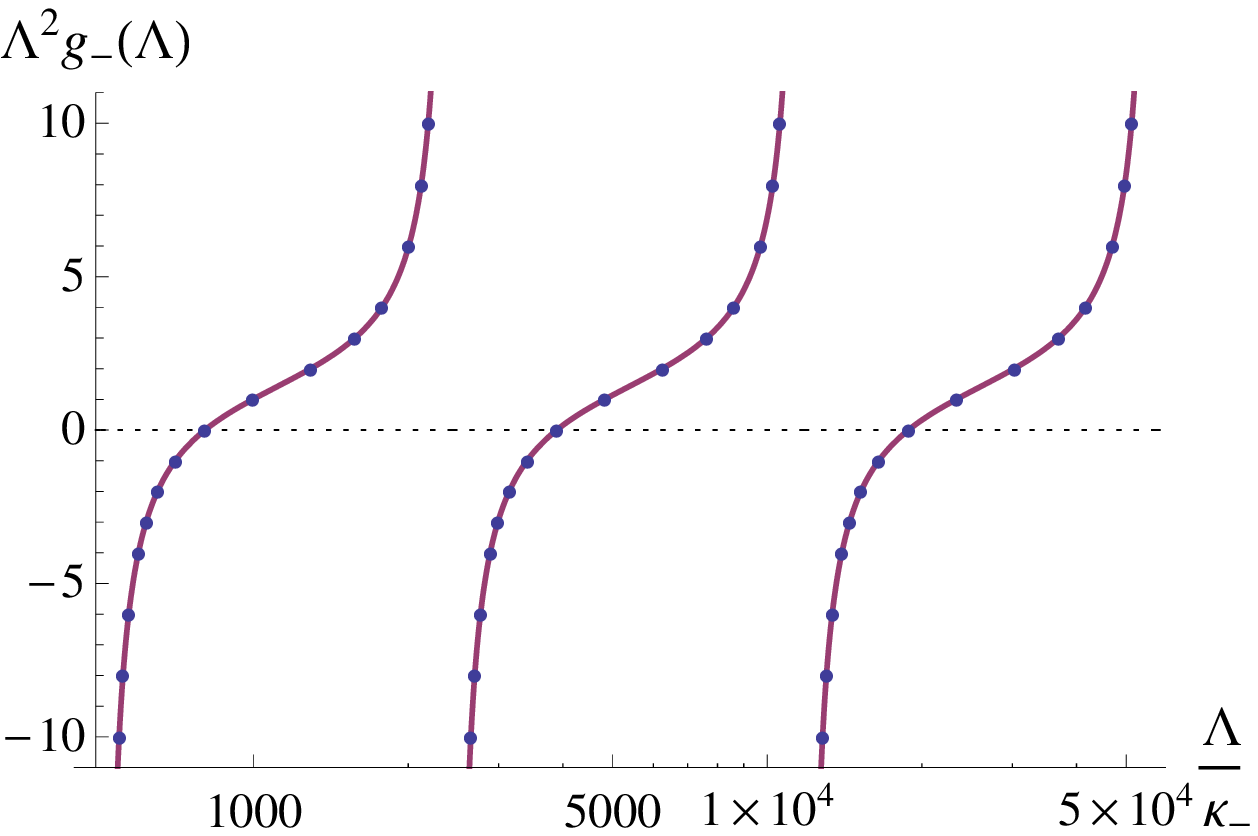}\hfill\hfill
 \caption{(Color online) Running couplings $g_+(\Lambda)$ (left) and
 $\Lambda^2g_-(\Lambda)$ (right panel) as functions of the cutoff
 $\Lambda/\kappa_\pm$ for $c=1$ with equal masses $m_1=m_2$.  Points are
 numerical solutions of the integral equation (\ref{eq:integral_eq2}),
 and solid curves are fits by the analytic expression (\ref{eq:RG}).
 \label{fig:limit_cycle}}
\end{figure*}

On the other hand, there exists another pattern of
discrete-scaling-symmetry violation which is common to bosons and
fermions.  For more than two particles, both even- and odd-parity
two-particle states contribute in general.  Because they obey
incommensurate scalings set by $\lambda_+$ and $\lambda_-$, the discrete
scale invariance breaks down.  However, this pattern does not take place
in the range $c>\pi$ where only the even-parity channel exhibits the
discrete scale invariance, and the odd-parity channel remains scale
invariant [see Eq.~(\ref{eq:s_1D})].  Here, it is possible that an
arbitrary number of fermions maintains the discrete scaling symmetry set
by $\lambda_+$.  Furthermore, when $c<0$, both even- and odd-parity
channels remain scale invariant and hence the whole system.
Accordingly, as shown in Fig.~\ref{fig:phase_diagram}, we expect
two-component fermions with equal masses to exhibit three phases with
distinct symmetries as a function of the inverse slope: phases with full
scale invariance at $c^{-1}<0$, discrete scale invariance at
$0<c^{-1}<1/\pi$, and no scale invariance at $1/\pi<c^{-1}$.  Their
many-body physics and phase transitions in between will be extremely
interesting and should be explored in the future.

\section{Effective field theory \label{sec:EFT}}

\subsection{Renormalization group limit cycle}
The scaling trap can be formulated in the language of effective field
theories.  A local-field theory to be considered is
\begin{subequations}\label{eq:field_theory}
 \begin{align}\label{eq:hamiltonian}
  H &= \int\!dx\!\left[-\psi_i^\+\frac{\nabla^2}{2m_i}\psi_i(x)
  - \frac1{\mu c|x|}\psi_1^\+\psi_2^\+\psi_2\psi_1(x)\right]
  \\[6pt] \label{eq:counterterm}
  & + \frac{g_+}{\mu}\psi_1^\+\psi_2^\+\psi_2\psi_1(0)
  + \frac{g_-}{\mu}\nabla[\psi_1^\+\psi_2^\+]\nabla[\psi_2\psi_1](0),
 \end{align}
\end{subequations}
where $i=1,2$ is summed and the space argument $(x)$ acts on all
operators on its left.  The first part [Eq.~(\ref{eq:hamiltonian})]
describes particles interacting by the zero-range interaction with the
linearly space-dependent scattering length.  As we discussed above, such
an interaction is singular at the origin and needs to be regularized by
introducing a cutoff.  The independence of physical quantities from the
cutoff is ensured by counterterms.  According to the above argument of
discrete-scaling-symmetry violation, $N$-body counterterms,
$\psi^{\+N}\psi^N(0)$ and $\nabla[\psi^{\+N}]\nabla[\psi^N](0)$, are
needed for bosons each with $N\geq2$.  On the other hand, the two
counterterms in Eq.~(\ref{eq:counterterm}) are expected to be sufficient
for an arbitrary number of two-component fermions with equal masses.

An expression for the cutoff-dependent coupling $g_\pm(\Lambda)$ is
obtained in the same spirit of Ref.~\cite{Bedaque:1999}.  We employ a
sharp momentum cutoff $|P|<\Lambda$ and require that two-body physics
becomes independent from the choice of $\Lambda$.  The two-body sector
of our field theory (\ref{eq:field_theory}) is equivalent to the
Schr\"odinger equation in Eq.~(\ref{eq:schrodinger}) but with the
modified interaction potential
\begin{equation}
 V(X,x) = \left[-\frac1{c|X|} + g_+\delta(X)
  + g_-\loarrow\nabla_X\delta(X)\roarrow\nabla_X\right]\frac{\delta(x)}{\mu}.
\end{equation}
Accordingly, the integral equation in Eq.~(\ref{eq:integral_eq}) is
modified to
\begin{equation}\label{eq:integral_eq2}
 \begin{split}
  \tilde\chi(P) &= -\frac1{\sqrt{P^2+\kappa^2}}\sqrt\frac{M}{\mu}
  \int_{-\Lambda}^\Lambda\!\frac{dP'}{2\pi} \\[4pt]
  & \times \left[\frac2c\ln\!\left(\frac{|P-P'|}\Lambda\right)
  + g_+(\Lambda) + g_-(\Lambda)PP'\right]\tilde\chi(P').
 \end{split}
\end{equation}
Then we require that its solution $\tilde\chi_\pm(P)$ for
$|P|,\kappa\ll\Lambda$ found in Eq.~(\ref{eq:solution}) does not change
when the cutoff is changed from $\Lambda$ to $\Lambda'$.  This
requirement is satisfied by choosing
\begin{subequations}\label{eq:RG}
 \begin{align}
  g_+(\Lambda) &= \frac2c\left\{\alpha_+
  - \frac{\cot[\phi_+(\Lambda)]}{s_+}\right\}, \\[8pt]
  \Lambda^2g_-(\Lambda) &= \frac2c\,\alpha_-
  \frac{s_-\cos[\phi_-(\Lambda)] + \sin[\phi_-(\Lambda)]}
  {s_-\cos[\phi_-(\Lambda)] - \sin[\phi_-(\Lambda)]},
 \end{align}
\end{subequations}
where
$\phi_\pm(\Lambda)\equiv s_\pm\ln(2\Lambda/\kappa_\pm)-\beta_\pm$.  With
numerical constants $\alpha_\pm$ and $\beta_\pm$, the analytic
expression (\ref{eq:RG}) fits to numerical solutions of the integral
equation (\ref{eq:integral_eq2}) accurately as shown in
Fig.~\ref{fig:limit_cycle}.\footnote{We found fitting parameters to be
$(\alpha_+,\beta_+)\approx(0.409,0.326),\,(0.256,0.237),\,(0.169,0.153)$
and
$(\alpha_-,\beta_-)\approx(1.35,0.220),\,(1.18,0.106),\,(1.11,0.0542)$
for $c\sqrt{4\mu/M}=0.5,\,1,\,1.5$, respectively.}
An important observation is that these couplings run in logarithmically
periodic ways in the large cutoff limit $\Lambda/\kappa_\pm\gg1$.  This
is known as a {\em renormalization group limit cycle\/}, and the Efimov
effect is its rare manifestation in physics~\cite{Wilson:2005}.  Our
scaling trap is newly added to a short list of systems demonstrating the
limit cycle.

\subsection{Universal relationships}
The field-theoretical formulation (\ref{eq:field_theory}) is useful to
derive universal relationships valid for any few-body or many-body
state~\cite{Braaten:2011}.  An operator product expansion of
\begin{align}\label{eq:OPE}
 & \int\!dx_1\,e^{-ip_1x_1}\,\psi_1^\+\!\left(X_1-\frac{x_1}2\right)
 \psi_1\!\left(X_1+\frac{x_1}2\right) \\[4pt]
 &= W_+(X_1,p_1)\O_+(X_1) + W_-(X_1,p_1)\O_-(X_1) + \cdots \notag
\end{align}
at $|p_1|\to\infty$ is dominated by the following two local operators:
\begin{subequations}
 \begin{align}
  \O_+(X) &\equiv \psi_1^\+\psi_2^\+\psi_2\psi_1(X), \\[8pt]
  \O_-(X) &\equiv \nabla[\psi_1^\+\psi_2^\+]\nabla[\psi_2\psi_1](X).
 \end{align}
\end{subequations}
By matching the matrix elements of both sides of Eq.~(\ref{eq:OPE}) with
respect to the two-particle trapped state obtained in
Sec.~\ref{sec:two-body}, the Wilson coefficient of $\O_\pm$ is found to
be
\begin{equation}
 W_\pm(X_1,p_1) = -\frac{M}{\mu}
  \frac{\d g_\pm}{\d\ln\kappa_\pm}t_\pm(p_1)\delta(X_1).
\end{equation}
Accordingly, the momentum distribution of a particle with mass $m_1$
exhibits the oscillatory large momentum tail
\begin{align}\label{eq:momentum}
 \rho(p_1) &= \int\!dX_1dx_1\,e^{-ip_1x_1}\left\<\psi_1^\+\!\left(X_1-\frac{x_1}2\right)
 \psi_1\!\left(X_1+\frac{x_1}2\right)\right\> \notag\\[4pt]
 &\to t_+(p_1)\,\C_++t_-(p_1)\,\C_-
\end{align}
for any state of the scaling trap.  The functional form of each term is
fixed by the two-body physics $t_\pm(p_1)$ obtained in
Eq.~(\ref{eq:tail}) for equal masses while its overall magnitude is set
by an analog of the contact density defined by
\begin{equation}
 \C_\pm \equiv -\frac{M}{\mu}
  \left\<\frac{\d g_\pm}{\d\ln\kappa_\pm}\O_\pm(0)\right\>.
\end{equation}
By applying the Hellmann-Feynman theorem to the Hamiltonian
(\ref{eq:field_theory}), we find that $\C_\pm$ measures how an energy of
the state under consideration changes with respect to $\kappa_\pm$:
\begin{equation}\label{eq:energy}
 \kappa_\pm\frac{\d E}{\d\kappa_\pm} = -\frac{\C_\pm}{M}.
\end{equation}
In particular, the contact density is given by
$\C_\pm=e^{-2\pi n/s_0}\kappa_\pm^2$ for a two-particle trapped state
with binding energy in Eq.~(\ref{eq:spectrum}).

Another universal relationship involving $\C_\pm$ can be derived by
considering a ``boundary'' operator product expansion of
\begin{equation}\label{eq:BOPE}
 \psi_1^\+\psi_2^\+\psi_2\psi_1(X)
  = W_+(X)\O_+(0) + W_-(X)\O_-(0) + \cdots
\end{equation}
toward the origin $X\to0$.  By matching the matrix elements of both
sides of Eq.~(\ref{eq:BOPE}) with respect to the two-particle trapped
state obtained in Sec.~\ref{sec:two-body}, the Wilson coefficient of
$\O_\pm$ is found to be
\begin{equation}
 W_\pm(X) = -\frac{M}{\mu}\frac{\d g_\pm}{\d\ln\kappa_\pm}u_\pm(X),
\end{equation}
where $u_\pm(X)\equiv\lim_{X\to0}|\psi_\pm(X,0)|^2/\kappa_\pm^2$ is
obtained from the two-body wave function in
Eq.~(\ref{eq:wave_function2}):
\begin{subequations}
 \begin{align}
  \begin{split}
   u_+(X) &= \biggl|N_+\frac{\cosh\frac{\pi s_+}2}{\pi}\Gamma(is_+) \\
   &\quad \times \cos\!\left[s_+\ln\!\left(\frac{\kappa_+|X|}2\right)
   - \arg\Gamma(is_+)\right]\biggr|^2,
  \end{split} \\[8pt]
  \begin{split}
   u_-(X) &= \biggl|N_-\frac{\sinh\frac{\pi s_-}2}{\pi}\Gamma(is_-) \\
   &\quad \times \cos\!\left[s_-\ln\!\left(\frac{\kappa_-|X|}2\right)
   - \arg\Gamma(is_-)\right]\biggr|^2.
  \end{split}
 \end{align}
\end{subequations}
Therefore, the probability of finding two particles at the same point
oscillates toward the origin as
\begin{equation}\label{eq:pair}
 \<\psi_1^\+\psi_2^\+\psi_2\psi_1(X)\>
  \to u_+(X)\,\C_+ + u_-(X)\,\C_-.
\end{equation}
These universal relationships found in Eqs.~(\ref{eq:momentum}),
(\ref{eq:energy}), and (\ref{eq:pair}) are valid for any few-body or
many-body state of the scaling trap.

\begin{figure*}[t]\hfill
 \includegraphics[width=0.9\columnwidth,clip]{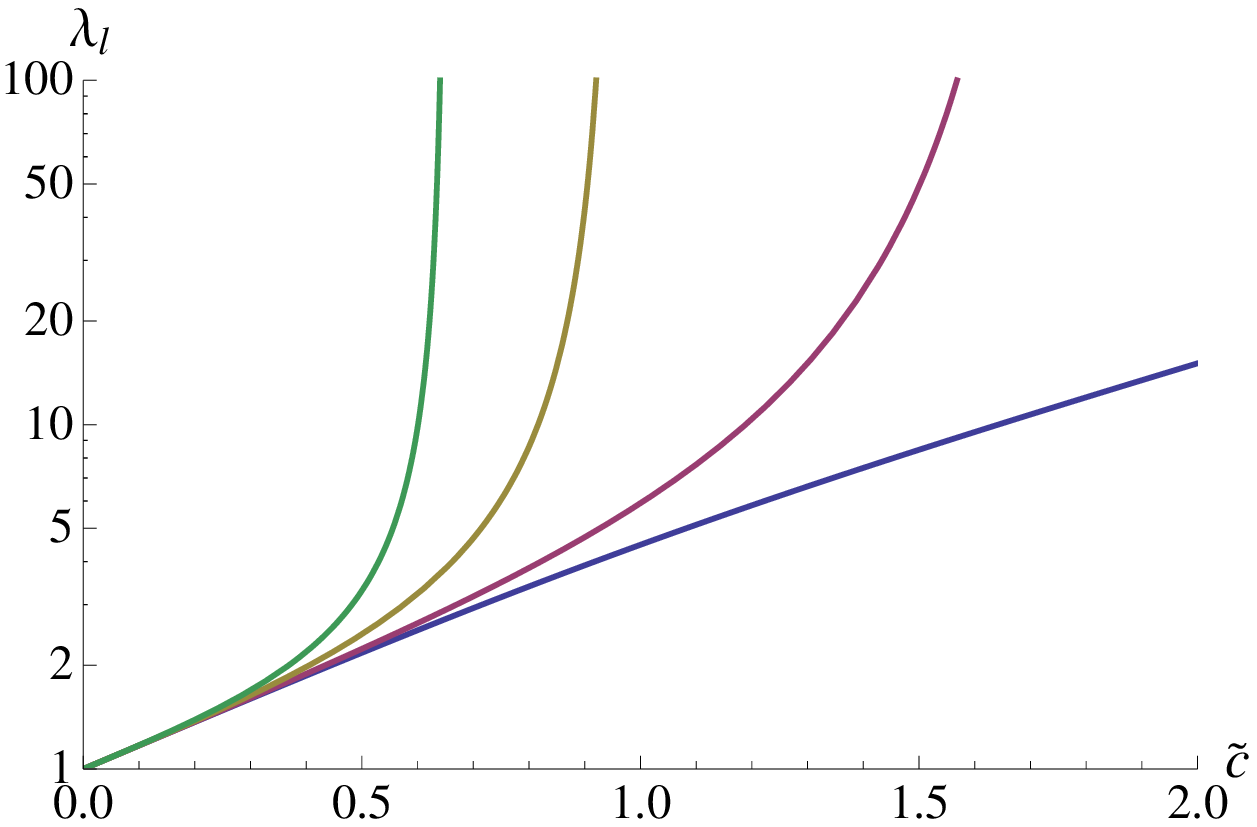}\hfill\hfill
 \includegraphics[width=0.9\columnwidth,clip]{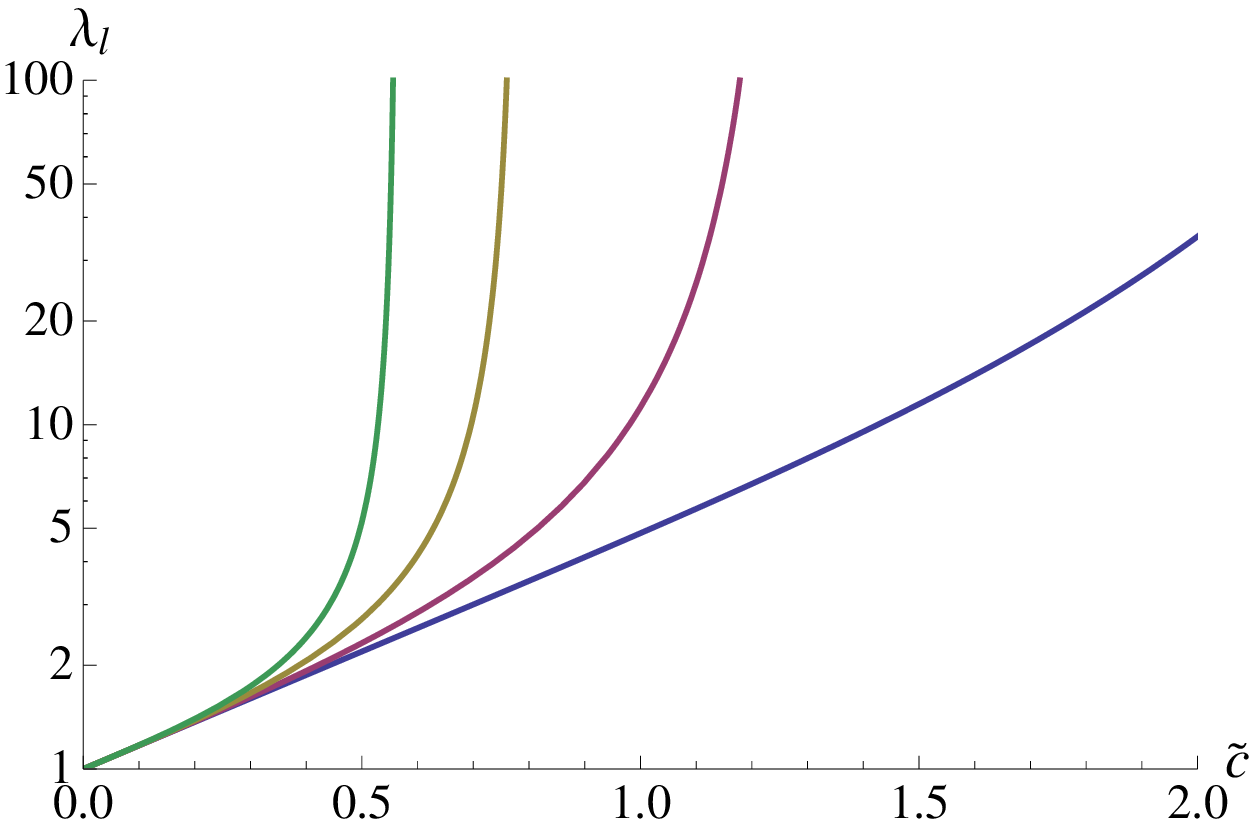}\hfill\hfill
 \caption{(Color online) Scale factor $\lambda_\ell=e^{\pi/s_\ell}$ in
 two (left) or three dimensions (right panel) determined by
 Eq.~(\ref{eq:s_2D}) or (\ref{eq:s_3D}) as a function of the slope of
 the scattering length $\tilde{c}\equiv c\sqrt{4\mu/M}$.  Four curves
 from the bottom to the top correspond to $\ell=0,\,1,\,2,\,3$,
 respectively.  \label{fig:scale-factor_2D-3D}}
\end{figure*}

\section{Scaling traps in two and three dimensions \label{sec:2D-3D}}
So far we have focused on physics in one dimension, but our idea works
equally well in two and three dimensions.  A two-body bound-state
problem by a zero-range interaction in an arbitrary spatial dimension
$d$ reduces to solving an integral equation which is an analog of
Eq.~(\ref{eq:integral_eq}) in real space:
\begin{equation}\label{eq:integral_eq_real}
 \chi_d(\X) = \int\!\frac{d\X'd\P}{(2\pi)^d}
  \frac{e^{i\P\cdot(\X-\X')}\,\chi_d(\X')}
  {\left[\sqrt{\frac{\mu}{M}}\sqrt{\P^2+\kappa^2}\,a(\X')\right]^{2-d}}.
\end{equation}
Here, $E=-\kappa^2/(2M)$ is the binding energy, and
\begin{equation}
 \chi_2(\X) \equiv \lim_{|\x|\to0}\frac{\psi(\X,\x)}{\ln[2e^{-\gamma}a(\X)/|\x|]}
\end{equation}
and
\begin{equation}
 \chi_3(\X) \equiv \lim_{|\x|\to0}\frac{\d}{\d|\x|}[|\x|\psi(\X,\x)]
\end{equation}
are associated with wave functions with two particles at the same
point.  The scattering length $a(\X)$ is defined so that when
$a(\X)=a>0$ is uniform, there is a single bound state with binding
energy $E=-1/(2\mu a^2)$.\footnote{Note that this scattering length in
$d=2$ is different from the conventional one defined by
$\lim_{|\x|\to0}\psi(\x)\propto\ln(a_*/|\x|)$ for which the binding
energy is given by $E=-4e^{-2\gamma}/(2\mu a_*^2)$.  Therefore,
$a_*=2e^{-\gamma}a$ in this paper, where $\gamma\approx0.5772$ is the
Euler-Mascheroni constant.}

The emergence of a scaling trap for the linearly space-dependent
scattering length $a(\X)=c|\X|$ is deduced from the existence of
solutions that oscillate as $\chi_d(\X)\to|\X|^{1-d\pm is}$ toward the
origin $\kappa|\X|\to0$.  By substituting a scaling ansatz
\begin{equation}
 \chi_2(\X) \sim |\X|^{-1+is_\ell}\,e^{i\ell\theta_{\hat\X}}
\end{equation}
or
\begin{equation}
 \chi_3(\X) \sim |\X|^{-2+is_\ell}P_\ell(\cos\theta_{\hat\X})
\end{equation}
into the integral equation (\ref{eq:integral_eq_real}) with
$\kappa\to0$, we find that $s_\ell$ solves
\begin{equation}\label{eq:s_2D}
 \ln\!\left(\frac1c\sqrt\frac{M}{4\mu}\right)
  = \frac{\Psi\!\left(\frac{\ell+1+is_\ell}2\right)
  +\Psi\!\left(\frac{\ell+1-is_\ell}2\right)}2
\end{equation}
in $d=2$ with $\Psi(z)\equiv\Gamma'(z)/\Gamma(z)$ or
\begin{equation}\label{eq:s_3D}
 \frac1c\sqrt\frac{M}{4\mu}
  = \frac{\Gamma\!\left(\frac{\ell+2+is_\ell}2\right)
  \Gamma\!\left(\frac{\ell+2-is_\ell}2\right)}
  {\Gamma\!\left(\frac{\ell+1+is_\ell}2\right)
  \Gamma\!\left(\frac{\ell+1-is_\ell}2\right)}
\end{equation}
in $d=3$.  For each angular momentum $\ell$ where $s_\ell$ has a
solution,\footnote{$s_\ell$ has a solution when
$c\sqrt\frac{4\mu}{M}<\exp\!\left[-\Psi\!\left(\frac{\ell+1}2\right)\right]\approx7.124,\,1.781,\,0.9642,\,0.6552,\,\ldots$
in $d=2$ or when
$c\sqrt\frac{4\mu}{M}<\left[\Gamma\!\left(\frac{\ell+1}2\right)/\Gamma\!\left(\frac{\ell+2}2\right)\right]^2\approx3.142,\,1.273,\,0.7854,\,0.5659,\,\ldots$
in $d=3$ for $\ell=0,\,1,\,2,\,3,\,\ldots\,$, respectively.}
there exists an infinite tower of two-particle trapped states
characterized by discrete scale invariance:
\begin{equation}
 E_\ell^{(n)} = -e^{-2\pi n/s_\ell}\,\frac{\kappa_\ell^2}{2M}.
\end{equation}
Its scale factor $\lambda_\ell\equiv e^{\pi/s_\ell}$ is plotted in
Fig.~\ref{fig:scale-factor_2D-3D}.  We expect similar patterns of
discrete-scaling-symmetry violation for more than two particles as
elucidated in one dimension.

\section{Remarks on experimental realization \label{sec:remarks}}
In this paper, we proposed scaling traps in which two particles form an
infinite tower of trapped states characterized by discrete scale
invariance.  The key idea is to make the scattering length proportional
to the distance from the center of the system so that a short-range
interaction becomes scale invariant.  Because of the lack of
translational symmetries, center-of-mass and relative motions of the
pair are coupled, and its relative motion generates an effective inverse
square potential for its center-of-mass motion.  Such a space-dependent
interaction can be realized in ultracold-atom experiments by a magnetic-
or optical-field-induced Feshbach resonance with spatially varying
magnetic- or optical-field intensity~\cite{Chin:2010} or, if the system
is confined in lower dimensions, by a confinement-induced resonance with
transverse confinement lengths varying along a longitudinal
direction~\cite{Olshanii:1998,Petrov:2001}.

If two particles correspond to different spin states of a fermionic
atom, our two-body bound states with Efimov character are long-lived
because three-body recombinations are strongly suppressed by the Pauli
exclusion principle~\cite{Petrov:2004}.  Furthermore, a scale factor can
be easily controlled by the slope of the scattering length.  This will
greatly facilitate an observation of the discrete scale invariance, for
example, by a radio-frequency spectroscopy or a time-of-flight
measurement.  Therefore, the scaling trap overcomes common difficulties
in ultracold-atom experiments of Efimov physics arising from an
instability of three-body bound states and their sizable scale factor
$\approx22.7$.  Advantages of the scaling trap over the Efimov effect
are summarized in Table~\ref{tab:comparison}.

\begin{table}[t]
 \caption{Comparison between the Efimov effect and the scaling trap in
 terms of (i) the dimensionality in which they can appear, (ii) the
 required number of particles, (iii) the stability of the bound states,
 (iv) the order at which the large momentum tail oscillates, and (v) the
 tunability of the scale factor $\lambda$.  \label{tab:comparison}}
 \begin{ruledtabular}
  \begin{tabular}{ccccccccc}
   & Conditions &&& Efimov Effect &&& Scaling Trap & \\\hline
   & (i) &&& 3D &&& 1D, 2D, 3D & \\
   & (ii) &&& 3 particles &&& 2 particles & \\
   & (iii) &&& unstable &&& long-lived & \\
   & (iv) &&& subleading &&& leading & \\
   & (v) &&& fixed &&& tunable & \\
  \end{tabular}
 \end{ruledtabular}
\end{table}

We also elucidated that the discrete scaling symmetry emergent for two
particles is inevitably violated for three or more bosons due to the
appearance of resonance states obeying different scaling laws while they
are absent for fermions.  It is possible that insights developed here
shed new light on Efimov physics.  Our scaling trap thus serves as a
tunable model system to investigate universal physics involving scale
invariance, quantum anomaly, and renormalization group limit cycle,
which are important in a broad range of quantum physics.

\acknowledgments
This work was supported by a LANL Oppenheimer Fellowship and the US
Department of Energy under Contract No.\ DE-FG02-03ER41260.

\appendix*
\section{Derivation of Eqs.~(\ref{eq:solution}) and (\ref{eq:s_1D})}
The integral equation (\ref{eq:integral_eq}) in the limit of an
infinitesimal cutoff $\epsilon\to0$ becomes
\begin{equation}\label{eq:integral_eq3}
 \sqrt{P^2+\kappa^2}\,\tilde\chi(P) = -\frac2c\sqrt\frac{M}{\mu}
  \int\!\frac{dP'}{2\pi}\ln(\epsilon|P-P'|)\,\tilde\chi(P').
\end{equation}
The analytic solution to this integral equation presented in
Eqs.~(\ref{eq:solution}) and (\ref{eq:s_1D}) can be derived in a way
similar to that used in Ref.~\cite{Gogolin:2008}.

In the even-parity channel
$\tilde\chi_+(P)\equiv[\tilde\chi(P)+\tilde\chi(-P)]/2$, the integral
equation (\ref{eq:integral_eq3}) in terms of a new variable
$P\equiv\kappa\sinh Q$ can be written as
\begin{align}\label{eq:integral_even}
 & \cosh Q\,\tilde\chi_+(Q) \\[2pt] \notag
 &= -\frac2c\sqrt\frac{M}{\mu}
 \int\!\frac{dQ'}{2\pi}\ln[\epsilon\kappa|\sinh(Q-Q')|]\cosh Q'\tilde\chi_+(Q').
\end{align}
By defining the Fourier transform of the integral kernel by
\begin{equation}
 \K_+(s) \equiv \int\!\frac{dQ}{2\pi}\,e^{iQs}\,\ln|\sinh Q|
  = -\frac{\coth\frac{\pi s}2}{2s},
\end{equation}
the integral equation (\ref{eq:integral_even}) can be brought into a
differential equation:
\begin{equation}
 \begin{split}
  & \cosh Q\,\tilde\chi_+(Q)
  = -\frac2c\sqrt\frac{M}{\mu}\,\K_+(i\d_Q)\cosh Q\tilde\chi_+(Q) \\
  &\qquad - \frac2c\sqrt\frac{M}{\mu}\,\ln(\epsilon\kappa)
  \int\!\frac{dQ'}{2\pi}\cosh Q'\tilde\chi_+(Q').
 \end{split}
\end{equation}
Since $\tilde\chi_+(Q)$ is an even function, its solution is easily
found to be
\begin{equation}
 \cosh Q\,\tilde\chi_+(Q) \propto \cos(s_+Q)
\end{equation}
with $s_+\neq0$ satisfying
\begin{equation}
 1 = -\frac2c\sqrt\frac{M}{\mu}\,\K_+(s_+).
\end{equation}

Similarly, in the odd-parity channel
$\tilde\chi_-(P)\equiv[\tilde\chi(P)-\tilde\chi(-P)]/2$, the integral
equation (\ref{eq:integral_eq3}) in terms of a new variable
$P\equiv\kappa\sinh Q$ can be written as
\begin{align}\label{eq:integral_odd}
 & \cosh Q\,\tilde\chi_-(Q) \\[2pt] \notag
 &= -\frac2c\sqrt\frac{M}{\mu}
 \int\!\frac{dQ'}{2\pi}\ln\left|\tanh\frac{Q-Q'}2\right|\cosh Q'\tilde\chi_-(Q').
\end{align}
By defining the Fourier transform of the integral kernel by
\begin{equation}
 \K_-(s) \equiv \int\!\frac{dQ}{2\pi}\,e^{iQs}\,\ln\left|\tanh\frac{Q}2\right|
  = -\frac{\tanh\frac{\pi s}2}{2s},
\end{equation}
the integral equation (\ref{eq:integral_odd}) can be brought into a
differential equation:
\begin{equation}
 \cosh Q\,\tilde\chi_-(Q)
  = -\frac2c\sqrt\frac{M}{\mu}\,\K_-(i\d_Q)\cosh Q\tilde\chi_-(Q).
\end{equation}
Since $\tilde\chi_-(Q)$ is an odd function, its solution is easily found
to be
\begin{equation}
 \cosh Q\,\tilde\chi_-(Q) \propto \sin(s_-Q)
\end{equation}
with $s_-$ satisfying
\begin{equation}
 1 = -\frac2c\sqrt\frac{M}{\mu}\,\K_-(s_-).
\end{equation}


\begin{thebibliography}{99}

\bibitem{Braaten:2006}
  E.~Braaten and H.-W.~Hammer,
  {\em Universality in few-body systems with large scattering length\/},
  Phys.\ Rept.\ {\bf 428}, 259 (2006).

\bibitem{Efimov:1970}
  V.~Efimov,
  {\em Energy levels arising from resonant two-body forces in a three-body system\/},
  Phys.\ Lett.\ B {\bf 33}, 563 (1970).

\bibitem{Ferlaino:2010}
  F.~Ferlaino and R.~Grimm,
  {\em Forty years of Efimov physics: How a bizarre prediction turned into a hot topic\/},
  Physics {\bf 3}, 9 (2010).

\bibitem{Nishida:2011} 
  Y.~Nishida and S.~Tan,
  {\em Liberating Efimov physics from three dimensions\/},
  Few-Body Syst.\ {\bf 51}, 191 (2011).

\bibitem{Landau-Lifshitz}
  L.~D.~Landau and E.~M.~Lifshitz, {\em Quantum Mechanics\/}
  (Butterworth-Heinemann, Oxford, 1977).

\bibitem{Holstein:1993}
  B.~R.~Holstein,
  {\em Anomalies for pedestrians\/},
  Am.\ J.\ Phys.\ {\bf 61}, 142 (1993).

\bibitem{Castin:2011}
  Y.~Castin and F.~Werner,
  {\em Single-particle momentum distribution of an Efimov trimer\/},
  Phys.\ Rev.\ A {\bf 83}, 063614 (2011).

\bibitem{Braaten:2011}
  E.~Braaten, D.~Kang, and L.~Platter,
  {\em Universal relations for identical bosons from three-body physics\/},
  Phys.\ Rev.\ Lett.\ {\bf 106}, 153005 (2011).

\bibitem{Deltuva:2012}
  A.~Deltuva,
  {\em Properties of universal bosonic tetramers\/},
  Few-Body Syst.\ (to be published)
  [arXiv:1202.0167 (physics.atom-ph)],
  and references therein.

\bibitem{Takahashi}
  M.~Takahashi,
  {\em Thermodynamics of One-Dimensional Solvable Models\/}
  (Cambridge University Press, Cambridge, 1999).

\bibitem{Bedaque:1999}
  P.~F.~Bedaque, H.-W.~Hammer, and U.~van~Kolck,
  {\em Renormalization of the three-body system with short range interactions\/},
  Phys.\ Rev.\ Lett.\ {\bf 82}, 463 (1999);
%
  {\em The Three boson system with short range interactions\/},
  Nucl.\ Phys.\ A {\bf 646}, 444 (1999).

\bibitem{Wilson:2005} 
  K.~G.~Wilson,
  {\em The origins of lattice gauge theory\/},
  Nucl.\ Phys.\ B (Proc.\ Suppl.) {\bf 140}, 3 (2005).

\bibitem{Chin:2010}
  C.~Chin, R.~Grimm, P.~Julienne, and E.~Tiesinga,
  {\em Feshbach resonances in ultracold gases\/},
  Rev.\ Mod.\ Phys.\ {\bf 82}, 1225 (2010).

\bibitem{Olshanii:1998}
  M.~Olshanii,
  {\em Atomic scattering in the presence of an external confinement and a gas of impenetrable bosons\/},
  Phys.\ Rev.\ Lett.\ {\bf 81}, 938 (1998).

\bibitem{Petrov:2001}
  D.~S.~Petrov and G.~V.~Shlyapnikov,
  {\em Interatomic collisions in a tightly confined Bose gas\/},
  Phys.\ Rev.\ A {\bf 64}, 012706 (2001).

\bibitem{Petrov:2004}
  D.~S.~Petrov, C.~Salomon, and G.~V.~Shlyapnikov,
  {\em Weakly bound dimers of fermionic atoms\/},
  Phys.\ Rev.\ Lett.\ {\bf 93}, 090404 (2004);
%
  {\em Scattering properties of weakly bound dimers of fermionic atoms\/},
  Phys.\ Rev.\ A {\bf 71}, 012708 (2005).

\bibitem{Gogolin:2008}
  A.~O.~Gogolin, C.~Mora, and R.~Egger,
  {\em Analytical solution of the bosonic three-body problem\/},
  Phys.\ Rev.\ Lett.\ {\bf 100}, 140404 (2008).

\end{thebibliography}
\end{document}